\documentstyle[12pt]{article}

\begin{document} 

\begin{center}  {\Large {\bf Autocorrelation of Hadrons in Jets
Produced in Heavy-Ion Collisions}}
\vskip .75cm
 {\bf Charles B. Chiu$^1$ and  Rudolph C. Hwa$^{2}$}
\vskip.5cm

 {$^1$Center for Particle Physics and Department of Physics\\ University of Texas at Austin, Austin, TX 78712-0264, USA\\
\bigskip
$^2$Institute of Theoretical Science and Department of
Physics\\ University of Oregon, Eugene, OR 97403-5203, USA}
\end{center}

\vskip.5cm

\begin{abstract} 
Autocorrelation of two pions produced in heavy-ion collisions at intermediate $p_T$ is calculated in the framework of the recombination model. The differences of the pseudo-rapidities and azimuthal angles of the two pions are related to the angle between two shower partons in a jet. It is shown how the autocorrelation distribution reveals the properties of jet cone of the shower partons created by high-$p_T$ partons in hard collisions.

\end{abstract}

\section{Introduction}

Correlations among hadrons produced at high and intermediate $p_T$
in heavy-ion collisions (HIC) have generated considerable interest in
their implications on how the hadronization mechanism and jet
structure may differ from those in $pp$ collisions \cite{ja}-\cite{mit}. 
Experimental investigations in the subjects can broadly be divided into
two types:  those that use triggers to identify near- and away-side jets
\cite{ja,dm,mit,pj} and those that use no triggers \cite{tr,mit,ja2,ja3}. Among the latter is the study of autocorrelations.  Autocorrelation
has been used extensively in time-series analysis. Its  application to multiparticle
production in HIC was pioneered by Trainor and his collaborators
\cite{tr,mit,tpp,lr}, and has generated a wealth of information
independent of triggers and their biases \cite{ja2,ja3}.  Theoretical
interpretation of autocorrelation has been slow in its development.  In
this paper we present the first prediction of how autocorrelation should
behave at  intermediate $p_T$ in the framework of parton
recombination \cite{hy}.

Theoretical studies of hadron correlation follow a wide variety of approaches; they differ mainly in the ways in which the interaction of jets with the medium is treated.  Energy loss of hard partons in medium has been extensively studied in pQCD \cite{bdm, wie, glv, wg, ww}, and its application to correlation on the same side of a trigger particle  has been considered with emphases placed on various separate but related issues, such as the medium-modified fragmentation functions, the effect of collective flow on jets, soft hadrons associated with medium-induced radiation, etc.\ \cite{whs, mww, asw, pp, wang, maj}. In such studies hadronization is usually treated by use of local parton-hadron duality (LPHD) or in terms of  fragmentation functions (FFs), neither of which can shed any light on the baryon puzzle in the intermediate $p_T$ region, e.g., the high $p/\pi$ ratio \cite{ppi}. Since the baryon problem has been well explained in the parton recombination model \cite{hy,hy1,du,tam}, the approach that we adopt here for the study of hadron correlations will be in the framework of that model. In that framework a number of investigations have already been carried out on dihadron correlations \cite{hy2, ht, rf, ch}. The medium effect  is taken into account  through the recombination of shower partons with the thermal partons that are in the environment, when the hard collision occurs near the surface of the medium. Our emphasis in this paper is on autocorrelation. Correlation of the hadrons produced on the side opposite to the trigger particle has also been investigated recently in numerous ways \cite{shu,kmw,dre,ren,vit}. Since the path length of a hard parton in the medium in such problems is large, they belong to a category of subjects not immediately relevant to our concern here, which is restricted to hadrons that are in the same jet.

The use of trigger has its advantages, especially in showing the
properties of the away-side azimuthal distribution that reveals the
effects of jet quenching.  However, it is necessary to subtract the
background, which is not unambiguous.  Autocorrelation is a measure
of the difference between two nearby values of a variable, with all other
variables being integrated over.  When those values are close, they are
dominated by contributions arising from the same jet in an event.  No
background subtraction is needed.  So far autocorrelation in the data from
the Relativistic Heavy-Ion Collider (RHIC) has been analyzed for
differences in pseudorapidity $\eta$ and in azimuthal angle $\phi$, but
only at low $p_T$ \cite{ja2,ja3}.  The model that we shall use to study
the autocorrelation in jets involve thermal-shower recombination for
which the reliable $p_T$ region is above 2 GeV/c.  Thus at this point
our predictions cannot be compared to the  results of the autocorrelation analysis of the experimental data. 
Nevertheless, on theoretical grounds it is of interest to show how the
angular distribution of shower partons can be related to the
autocorrelation of pions in the differences in $\eta$ and $\phi$. We await with anticipation the relevant data that are forthcoming.

\section{The Problem}

Autocorrelation is a measure of two-particle correlation in multiparticle
production with a minimum loss of information and without trigger
bias.  It has been studied extensively in $pp$ and HIC \cite{ja2,ja3,tpp}.
At intermediate to high $p_T$ the dominant contribution to the
two-particle correlation comes from jets, when the momenta for the two
particles are close together.  Our problem in this paper is to calculate the
autocorrelation distribution from a model in which the shower partons in
a jet has certain prescribed properties, and in this section we outline our
plan of attack, leaving the details to the following sections.

Let $x_i$ be an attribute of the momentum $\vec{p}_i$ of the {\it i} th
particle, such as its angle relative to some axis.  The two-particle
correlation function in terms of $x_i$ is
\begin{eqnarray} 
C_2 (x_1, x_2) = \rho_2 (x_1, x_2) -  \rho_1 (x_1) \rho_1( x_2)
\label{1}
\end{eqnarray}
where $\rho_1$ and $\rho_2$ are one- and two-particle distributions,
respectively.  For autocorrelation one defines the sum and difference of
$x_i$
\begin{eqnarray}
x_{\pm} = x_2 \pm x_1
\label{2}
\end{eqnarray}
and rewrite Eq.\ (\ref{1}) in the form
\begin{eqnarray} 
C_2 (x_+, x_-) = \rho_2 (x_+, x_-) -  \rho_1 ((x_+-x_-)/2) \rho_1((x_++x_-)/2) \ .
\label{3}
\end{eqnarray}
This would vanish if there is no correlation in $\rho _2$.  Anticipating
correlation in the variable $x_- $ and mild dependence on $x_+$, one
defines the autocorrelation distribution to be 
\begin{eqnarray}
A(x_-) = {1\over R}\int_R dx_+ C_2 (x_+, x_-) \ ,
\label{4}
\end{eqnarray}
where the integration is carried out over a range $R$.
Usually, if there are other variables in the problem, which are, however,
not germane to the correlation measure, they are integrated over also. 
If the range $R$ is wide enough so that the boundaries depend on $x_-$, then one must proceed carefully to account for the $x_-$ dependence arising from $R$. Details of that problem related to binning can be found in \cite{tpp,tt}. In our consideration in the following $R$ will be small enough not to involve such complications.
Clearly, only the correlated part in $\rho_2 (x_+, x_-)$ contributes to
$A(x_-)$.  If $\rho_2 (x_+, x_-)$ has mild dependence on $x_+$, not
much information is lost by the integration in Eq.\ (\ref{4}).  $A(x_-)$
treats the two particles on equal footing and requires no subtraction of
background besides what is explicit in Eq.\ (\ref{3}).

In HIC if the transverse components of $\vec{p}_i$ are $>2$ GeV/c,
then jets are involved; furthermore, if the angle between $\vec{p}_1$ and $\vec{p}_2$ is less than, say, $\pi/4$, then the two particles are highly likely to be the
particles in the same jet.  Thus the angular differences between the
momentum vectors provide information about the structure of the jet. 
Let the angular variables of $\vec{p}_1$ and $\vec{p}_2$, referred to
the longitudinal axis, be $(\theta_1,\phi _1)$ and $(\theta_2,\phi _2)$,
respectively.  Define
\begin{eqnarray}
\theta_{\pm} = \theta_2 \pm \theta_1 , \quad \phi_{\pm} = \phi_2
\pm \phi_1 .
\label{5}
\end{eqnarray}
In a central collision $\phi _+$ is irrelevant by azimuthal symmetry,
which we shall assume.  Thus the essential variables for correlation are
$\theta_+, \theta_-$ and $\phi_-$.  If the correlation function is
determined experimentally at midrapidity with a narrow rapidity
window, the range of  $ \theta_+$ is not
large.  Then in applying Eq.\ (\ref{4}) to this problem, we have
\begin{eqnarray}
A (\theta_-, \phi_-) ={1\over R_{\theta_+}} \int_{R_{\theta_+}} d\theta _+ C_2 (\theta_+, \theta_-, \phi_-) \ ,
\label{6}
\end{eqnarray} 
where $R_{\theta_+}$ is some range in $\theta_+$ that is not constrained by the ranges of $\theta_-$ and $\phi_-$ of interest.
Experimentally, the $p_T$ variables are integrated over specific ranges
of choice, and are not expressed explicitly.  It is important to note that
$A (\theta_-, \phi_-)$ depends only on the difference angular variables,
and is therefore, in principle, independent of the coordinate system in
which the angles are defined.  The only angle associated with two
momentum vectors that is independent of the coordinate system is the
angle $\chi$ between the two vectors, i.e., 
\begin{eqnarray}
\cos \chi = \cos \theta_1 \cos \theta _2 + \sin \theta_1\sin \theta _2
\cos (\phi _2 - \phi_1) \quad ,
\label{7}
\end{eqnarray}
which can be expressed in terms of $\theta _{\pm}$ and $\phi _-$ as 
\begin{eqnarray}
\cos \chi = {1 \over 2}\left[\cos \theta_- (1 +\cos \phi _-) + \cos
\theta_+ (1 -\cos \phi _-)  \right] \ .
\label{8}
\end{eqnarray}
How this is used in Eq.\ (\ref{6}) to obtain the autocorrelation
distribution will be discussed in Sec. 4.

The angle $\chi$ provides the crucial link between the observables and
the variables that can suitably describe the dynamics of the partons that
are associated with jets.  At intermediate and high $p_T$ in HIC the
partons that hadronize are the shower partons in jets produced by hard
scattering.  The axis in reference to which those partons can best be
described is the jet axis.  For detection at midrapidity the jet directions
are approximately perpendicular to the beam axis.  Thus to relate the
angular variables of the partons, referred to the jet axis, to the angular
variables of the hadrons, referred to the beam axis, is a complicated
geometrical problem.  In the recombination model the hadrons and the
constituent partons are collinear.  The angle between two shower
partons is therefore also the angle between the two pions that they
hadronize into.  That angle is $\chi$.  Hence, $\chi$ serves as the bridge
that connects the momentum space of the partons in the underlying
dynamics to the momentum space of the hadrons that can be
measured.  Autocorrelation in terms of $\theta_-$ and $\phi_-$ can
thus directly reveal the angular properties of the shower partons via
$\chi$.  To put this strategy into concrete formulation, we proceed first
to the momentum space of the partons in the next section, and then to
the momentum space of the hadrons in the following section.

\section{Correlation between Shower Partons}

Suppose that in a HIC a hard scattering takes place that sends a parton
to a momentum $\vec{k}$ with a large transverse momentum.  Such a
hard parton generates a shower of partons whose momentum-fraction
distributions have been determined in Ref.\ \cite{hy3}.  
It should be stressed that the shower partons under discussion here are not to be identified with the radiated gluons in pQCD. Those shower partons are defined by their recombination with each other to form hadrons and their distributions are determined by fitting the appropriate non-perturbative FFs in the recombination formalism \cite{hy3}. The gluons emitted by a hard parton, on the other hand, are the products of an evolution process well formulated in pQCD, but their hadronization is not well treated because the process is not perturbative. Of course, the physics of the shower partons that recombine and the gluons that are perturbatively radiated are related in some non-perturbative way, which need not distract us here. Since our goal is to determine the correlation between hadrons at intermediate $p_T$, we consider the shower partons that are defined for immediate hadronization by recombination in that $p_T$ range.

As stated above, those shower partons recombine among themselves to form hadrons whose momentum-fraction distributions are the FFs of the initiating hard parton. But when the hard parton is produced near the surface of a thermal medium, the shower partons can also recombine with the soft thermal partons in the environment to form hadrons at intermediate $p_T$.  In the recombination model that is the way how hadron production is affected by the medium.  Thus for pion production one considers $\cal ST$ recombination, and for proton production one has $\cal STT$ and $\cal SST$ recombination, where $\cal{S}$ denotes the shower-parton
distribution (SPD) and $\cal{T}$ the thermal parton distribution. Indeed, it has been shown in such an approach that the $p/\pi$ ratio in Au+Au collisions can be as large as 1 at $p_T\sim 3$ GeV/c \cite{hy}.

Now to study the correlation between two pions, both in the intermediate $p_T$ range, we need to consider two shower and two thermal partons in the combination $(\cal{ST})(\cal{ST})$, where the parentheses in $(\cal{ST})$ denote the partners that are to recombine.  Since only collinear partons can
recombine, the angle $\chi$ between the two pions that are formed is the 
same as the angle between the two
shower partons, whose momenta magnitudes are generally larger than
the two softer thermal partons.  Let us then concentrate on the two
shower partons and consider the joint distribution
\begin{eqnarray}
\left\{ \cal{SS}\right\}^{jj^{\prime}} \left(\vec{q}_1,  \vec{q}_2\right)
= \xi \sum_i \int dk kf_i (k) \left\{ S^j_i\left({q_1  \over  k},
\psi_1, \beta_1\right), S^{j^{\prime}}_i\left({q_2  \over 
k-q_1},
\psi_2, \beta_2\right)
\right\} ,
\label{9}
\end{eqnarray}
which is in the form that has been used previously in Ref. \cite{ch}.  The
curly brackets denote symmetrization of the momentum fractions of the
two shower partons \cite{hy}, but it is a process that is not important in
the following, since our emphasis will be on the angular variables.  For
the same reason, we shall not be concerned with $\xi$, which denotes
the fraction of hard partons that emerge from the dense medium to
hadronize, and the integral over $k$, weighted by $f_i (k)$ that is the
distribution of the hard parton $i$ in HIC.  $\psi_1$ and $\psi_2$ are
the polar angles of the $j$ and $j^{\prime}$ shower parton momenta,
$\vec{q}_1$ and $\vec{q}_2$, with reference to the jet axis, $\hat{k}$;
$\beta_1$ and $\beta_2$ are their corresponding azimuthal angles.
$S_i^{j,j^{\prime}}$ are the SPDs.  

The angle between $\vec{q}_1$ and $\vec{q}_2$, denoted by $\chi$
also, can be expressed similarly as in Eq.\ (\ref{7})
\begin{eqnarray}
\cos \chi = \cos \psi_1 \cos \psi_2 + \sin \psi_1 \sin \psi_2 \cos
(\beta_2 - \beta _1)\ .
\label{10}
\end{eqnarray}
If we define, as before,
\begin{eqnarray}
\psi_{\pm} = \psi _2 \pm \psi_1, \qquad \beta _- = \beta_2 - \beta_1,
\label{11}
\end{eqnarray}
then we have the alternate form similar to Eq.\ (\ref{8}), which in turn
can be written as
\begin{eqnarray}
\cos \chi  = A + B \cos \beta ,
\label{12}
\end{eqnarray}
where
\begin{eqnarray}
A = {1 \over 2 } (\cos \psi_- + \cos \psi_+),
\label{13}
\end{eqnarray}
\begin{eqnarray}
B = {1 \over 2 } (\cos \psi_- - \cos \psi_+),
\label{14}
\end{eqnarray}
and $\beta = \beta_-$ for brevity.  Note that $A$ and $B$ are linear in
$\cos \psi_{\pm}$, while the corresponding terms in Eq.\ (\ref{10}) are
quadratic in $\cos \psi_{1 , 2}$.  Since $\chi$ serves as the bridge to
the hadron momentum space, we want to determine here the
distribution in $\chi$ that correspond to the dynamical properties of
the jet cone defined with respect to $\vec{k}$.

The joint shower-parton distribution, as expressed in Eq.\ (\ref{9}), is
based on the assumption that the shower partons are dynamically
independent, but kinematically constrained by momentum
conservation
\begin{eqnarray}
q_1 + q_2 \leq k .
\label{15}
\end{eqnarray}
The constraint on the momentum magnitudes in Eq.\ (\ref{15}) does not imply angular constraint on $\psi_1$ and $\psi_2$. We shall assume  angular independence
 of the two partons, which  implies that their
distribution $G_2 (\psi_1,\psi_2)$ is factorizable
\begin{eqnarray}
G_2 (\psi_1,\psi_2) = G_1 (\psi_1) G_1 (\psi_2),
\label{16}
\end{eqnarray}
where $G_1(\psi)$ is the single-parton angular distribution in a jet cone. 
We assume that it has a Gaussian form
\begin{eqnarray}
G_1(\psi) = \exp \left(-\psi^2/2\sigma^2 \right)
\label{17}
\end{eqnarray}
with a width $\sigma$ that is basically unknown, since no reliable
theory is calculable at intermediate $k$.  We make the assumption here that $\sigma$ is independent of the momentum fraction of the shower parton in a jet. That assumption renders significant simplification of our consideration below, and is a reasonable first step in this exploratory study of autocorrelation.
If we can relate $\sigma$ to some observable through autocorrelation, we will
have achieved the objective of probing the microscopic dynamics by
phenomenology, at least in the first approximation.

As stated in the beginning of this section, the shower partons considered in Eq.\ (\ref{16}) are not directly identifiable as the gluons and their associated low-virtuality partons emitted by the hard parton in pQCD because the properties of the partons $j$ and $j'$ in Eq.\ (\ref{9}) are not determined by any perturbative calculation, but by fitting the non-perturbative FFs through recombination. Although it is known that a branching process can lead to angular ordering of the emitted partons \cite{ow, fw}, the physics of which has recently been applied to the study of jets traversing dense medium \cite{bw}, the formalism is  more reliable at higher jet energy than what we are concerned with here at intermediated $p_T$. Besides, hadronization by LPHD is what we can avoid when recombination has the advantage of proven phenomenological reliability. Note also that we study hadron correlation in jets produced near the surface of a dense medium, so the initiating hard partons do not have long path lengths through the medium. The only medium effect that we consider is the recombination with thermal partons, which does not, in first approximation, influence the angular consideration to follow.
For these various reasons the angles $\psi_1$ and $\psi_2$ in Eq.\ (\ref{16}) cannot be assigned the angular ordering property of the emitted partons in pQCD. For a simple first try we assume factorizability in Eq.\ (\ref{16}) and the independence of $\sigma$ on the momentum fraction of the shower parton. In an extension of this study in the future it would be reasonable to investigate the scenario where $\sigma$ increases with decreasing momentum fraction. But for constant $\sigma$ considered here the momentum-dependent part factors out and the angular properties are exposed with transparency. 

Given the two-parton angular distribution $G_2 (\psi_1,\psi_2)$ shown
in Eqs.\ (\ref{16}) and (\ref{17}) we can calculate the $\chi$
distribution by allowing for all possible orientations of the two vectors
$\vec{q_1}$  and $\vec{q_2}$.  Denoting it by $H(\chi)$, we have
\begin{eqnarray}
H(\chi) = \int d\cos \psi_2\, d\cos \psi_1\, d\beta\, G_2 (\psi_1, \psi_2)\,
\delta\left[\cos \chi - (A + B \cos \beta)\right],
\label{18}
\end{eqnarray}
where Eq.\ (\ref{12}) has been used to constrain the four angles
$\psi_{\pm}, \beta$ and $\chi$.  Since $H(\chi)$ is invariant under the
interchange of $\vec{q_1}$  and $\vec{q_2}$, we consider only the ranges
\begin{eqnarray}
0 \leq \psi_1 \leq \psi_2 \leq \alpha  .
\label{19}
\end{eqnarray}
In actual calculation we set $\alpha = \pi /4$.  In view of the definitions
of $A$ and $B$ given in Eqs.\  (\ref{13}) and (\ref{14}), we change the
integration variables to $\psi_{\pm}$ and $\beta$.  After some algebra
we obtain
\begin{eqnarray}
H(\chi) = {1 \over 4} \int ^{2\alpha} _{\chi} d \psi _+ \int
^{\psi_m}_0 d \psi _-{(\cos \psi_- - \cos \psi_+) g(\psi_+, \psi_-)
\over 
\left[(\cos \psi_- - \cos \chi)(\cos \chi - \cos \psi_+)
\right]^{1/2}}
\label{20}
\end{eqnarray}
where
\begin{eqnarray}
\psi_m = \min (\chi, 2 \alpha - \psi_+), 
\label{21}
\end{eqnarray}
and
\begin{eqnarray}
g (\psi_+, \psi_-) = \exp \left(-{\psi^2_+ + \psi^2_-  \over 4 \sigma^2
}\right) .
\label{22}
\end{eqnarray}
In Fig.\ 1 (a) we show $H(\chi)$ for four values of $\sigma$. To see more clearly the dependence of the width of $H(\chi)$ on the width $\sigma$ of $G_1(\psi)$, we normalize the function $H(\chi)$ by its value at $\chi=0$, by defining
\begin{eqnarray}
\tilde H(\chi)={H(\chi) \over H(0)}.   \label{22a}
\end{eqnarray}
It can be shown that in the approximation $\sigma \ll \alpha, H(\chi)$ has the limit, as $\chi\to 0$, $H(0)\approx (\pi/4)\sigma^2\exp (-\sigma^2/4)$. The corresponding normalized distribution $\tilde H(\chi)$ is  shown in Fig.\ 1 (b). In the $\sigma$ range illustrated,  the width of $\tilde{H}(\chi)$ is very closely  given by $\sigma_H=\sqrt{2}\sigma$.

It is important to recognize that the factorizable form of $G_2(\psi_1, \psi_2)$ in Eq.\ (\ref{16}) does not imply an absence of correlation between $\psi_1$ and $\psi_2$. That is because the two shower partons with momenta $\vec q_1$ and $\vec q_2$ are from the same jet, and the angles $\psi_1$ and $\psi_2$ refer to the same jet axis along $\vec k$. When $G_2(\psi_1, \psi_2)$ is included in the expression for $\{{\cal SS}\}^{jj'}(\vec q_1, \vec q_2)$ in Eq.\ (\ref{9}), the contribution to the two-parton distribution $\rho_2(\vec q_1, \vec q_2)$ is not factorizable. On the other hand, $\rho_2(\vec q_1, \vec q_2)$ contains a factorizable part $\rho_1(\vec q_1)\rho_1(\vec q_2)$, in which the two shower partons belong to two independent jets that are randomly related to each other. Such a component would be cancelled in the calculation of the correlation function $C_2(\vec q_1, \vec q_2)$ that can be defined as in Eq.\ (\ref{1}) for partons. 

Finally, before leaving this section on parton correlation it should be noted that despite their appearances the peaks in Fig.\ 1 should not be confused with the common notion of the properties of jet cone. $H(\chi)$ is the distribution of $\chi$, the angle between $\vec q_1$ and $\vec q_2$, neither of which need to be close to the jet axis. That is, $\chi$ should not be interpreted as the angle between the hard parton and a radiated parton. A collection of all events with a fixed value of $\chi$ is a large set that contains a small subset in which a parton is radiated at an angle $\chi$ with respect to the jet axis.
This point is so important that a simple example might be useful to make clear the difference. Suppose $\vec q_1$ and $\vec q_2$ are collinear so that $\chi=0$. However, both $\vec q_1$ and $\vec q_2$ can be directed at some angle $\psi_1=\psi_2$ that need not be zero, so that $\chi$ can in no way represent the angle between a radiated parton and the jet axis. Thus $H(\chi)$ should not be taken as a representation of the angular distribution of a radiated parton relative to the initiating hard parton. Since $\chi$ will play a principal role in autocorrelation,  this distinction underlies the difference between autocorrelation and the correlation based on triggers.

\section{Autocorrelation between Hadrons}

Having obtained the distribution in $\chi$, the angle between two shower partons in a jet, we can now determine the correlation between
two hadrons in a jet. We shall consider only the pions, since they are the dominant hadrons in a jet. As we have stated earlier, we consider the intermediate $p_T$ region where the thermal-shower recombination is most important, i.e., $3<p_T<8$ GeV/c in central Au+Au collisions \cite{hy, hy2, ch}. 
Furthermore, we restrict our attention here to only the pions produced at midrapidity so that all momentum vectors are nearly transverse to the beam direction. In  that case we can simplify our notation by denoting the pion transverse momentum $p_T$ and the parton transverse momentum $q_T$ by $p$ and $q$, respectively.
The $\cal (TS)(TS)$ contribution to the two-particle distribution is then
\begin{eqnarray}
{dN_{\pi\pi}^{TSTS}\over p_1dp_1d\eta_1d\phi_1p_2dp_2d\eta_2d\phi_2}={1\over (p_1p_2)^3}\int dq_1dq_2d\cos\psi_1d\cos\psi_2d\beta_1d\beta_2{\cal T}(p_1-q_1){\cal T}(p_2-q_2) \nonumber \\
\{{\cal SS}\}(q_1,\psi_1,\beta_1;q_2,\psi_2,\beta_2)\Delta(\psi_1,\beta_1,\eta_1,\phi_1;\psi_2,\beta_2,\eta_2,\phi_2),  \label{23}
\end{eqnarray}
where $\cal T$ is the thermal distribution of exponential form \cite{hy}, and $\cal \{SS\}$ is given in Eq.\ (\ref{9}). The recombination functions, written out explicitly in Ref. \cite{ch}, have already been integrated over, resulting in (a) the momenta of the thermal partons being $p_i-q_i$, and (b) a constraint between the pions' angular variables $\theta_i,\phi_i$ and the shower partons' angular variables $\psi_i, \beta_i$, contained in the $\Delta$ function in Eq.\ (\ref{23}). 
We have exhibited only the essential content of the $\cal (TS)(TS)$ recombination, where the momenta of the thermal and shower partons that recombine are collinear, so all the dynamical characteristics of the problem are in $\cal \{SS\}$, which we have studied in the preceding section, while all the kinematical relationships to the observed pions are in $\Delta$. Although the pseudorapidity variables $\eta_1$ and $\eta_2$ appear in Eq. (\ref{23}), it is only a technical step to be taken later to relate their difference to the angular differences, $\theta_-$ and $\phi_-$, that appear in the autocorrelation, Eq.\ (\ref{6}). Our task now is to first relate $A(\theta_-, \phi_-)$ to the two-particle distribution in Eq.\ (\ref{23}).

In our study of the autocorrelation our emphasis has been on the dependence on the angular differences, $\theta_-$ and $\phi_-$, relegating the $p_T$ values to the category of other variables that are to be integrated over. If we define the full correlation function to be 
\begin{eqnarray}
C_2(1,2)=\rho_2(1,2)-\rho_1(1)\rho_1(2),   \label{24}
\end{eqnarray}
where the arguments symbolize the whole sets of variables of the two detected particles, then the LHS of Eq.\ (\ref{23}) represents the $TSTS$ component of $\rho_2(1,2)$, which we now denote by $\rho_2^{TSTS}(1,2)$. Both in the analysis of the experimental data and in our construction of autocorrelation, the $p_T$ values are integrated over some chosen ranges. In our work here the autocorrelation does not depend sensitively on that $p_T$ range so long as it is the intermediate $p_T$ range where $TS$ recombination is dominant for the formation of a pion. Let us then define the integrated correlation
\begin{eqnarray}
\Gamma(\eta_1,\phi_1,\eta_2,\phi_2)=\int dp_1dp_2\, p_1p_2\, C_2(1,2).  \label{25}
\end{eqnarray}
With $\eta_\pm$ defined by 
\begin{eqnarray}
\eta_\pm=\eta_2\pm\eta_1     \label{26}
\end{eqnarray}
and $\phi_\pm$ given in Eq.\ (\ref{5}), $\Gamma(\eta_\pm, \phi_\pm)$ receives its contribution mainly from the $TSTS$ component, $\rho_2^{TSTS}(1,2)$, expressed in Eq.\ (\ref{23}), when $\eta_-$ and $\phi_-$ become small.

We now focus on  $\rho_2^{TSTS}(1,2)$ and, in particular, on the $\Delta$ function in Eq.\ (\ref{23}), which contains only the angular variables of the parton and pion momenta. As we have emphasized earlier, with the autocorrelation $A(\theta_-,\phi_-)$ as the aim of our analysis, the angle $\chi$ is the bridge between the parton and pion angular variables that are independent of the coordinate system. With that simplification in mind we write $\Delta$ in the form
\begin{eqnarray}
\Delta(\psi_1,\beta_1,\eta_1,\phi_1;\psi_2,\beta_2,\eta_2,\phi_2)= \hskip6cm  \nonumber \\
\int d\cos\chi\, \delta[\cos\chi-(A+B\cos\beta_-)]\ 
 \delta[\cos\chi-(C+D\cos\phi_-)],  \label{27}
\end{eqnarray}
where $A$ and $B$ are functions of $\psi_\pm$, defined in Eqs. (\ref{13}) and (\ref{14}), while $C$ and $D$ are similar functions of $\theta_\pm$ that follow from Eq.\ (\ref{8}), and are explicitly
\begin{eqnarray}
C={1\over 2}(\cos\theta_- + \cos\theta_+),  \label{28}\\
D={1\over 2}(\cos\theta_- - \cos\theta_+).  \label{29}
\end{eqnarray}
It is clear that the two $\delta$ functions in Eq. (\ref{27}) express the two ways of referring $\chi$ to the two different sets of angles defined with respect to the jet axis on the one hand, and to the beam axis on the other.

Substituting Eqs.\ (\ref{9}) and (\ref{27}) into Eq.\ (\ref{23}), and performing the angular integration as in Eq.\ (\ref{18}), we obtain $H(\chi)$ multiplied by some factors involving the momenta magnitudes. Such factors are irrelevant to our study, so we normalize it out by defining 
\begin{eqnarray}
\tilde C_2(\eta_\pm, \phi_\pm) = { \Gamma(\eta_+,\eta_-;\phi_+,\phi_-)\over \Gamma(\eta_+,0;\phi_+,0)}.  \label{30}
\end{eqnarray}
Pending the connection between $\eta_\pm$ and $\theta_\pm$, we have from the above process of integrations
\begin{eqnarray}
\tilde C_2(\theta_\pm, \phi_\pm)=\int d\cos\chi\, \tilde H(\chi)\, \delta[\cos\chi -(C+D\cos\phi_-)].   \label{31}
\end{eqnarray}
This integration over $\chi$ need not be performed explicitly, since it is only necessary to determine the domain of $\theta_\pm$ and $\phi_-$ that corresponds to each fixed value of $\chi$ and denote the result as $\tilde H(\theta_\pm, \phi_-)$. The final step is to integrate over $\theta_+$ as in Eq.\ (\ref{6}). Conceptually, it is simpler to visualize the average angle  $\bar\theta$, defined by
\begin{eqnarray}
\bar\theta=\theta_+/2,   \label{32}
\end{eqnarray}
which should be around $\pi/2$, since the two detected pions are restricted to a narrow interval at mid-rapidity, corresponding to a jet at roughly $\pi/2$ relative to the beam axis. If we denote the interval of $\bar\theta$ by $2\epsilon$, then we have
\begin{eqnarray}
\tilde A(\theta_-, \phi_-)={A(\theta_-, \phi_-)\over A(0, 0)}={1\over 2\epsilon}\int_{\pi/2-\epsilon}^{\pi/2+\epsilon} d\bar\theta\, \tilde C_2(\theta_+, \theta_-, \phi_-),   \label{33}
\end{eqnarray}
while $\phi_+$ is an irrelevant angle in an azimuthally symmetric problem. Experimentally, $\tilde A(\theta_-, \phi_-)$ is to be determined by use of Eqs.\ (\ref{6}) and (\ref{24}).

The transference from $\theta_-$ to $\eta_-$ involves $\theta_+$ and is just an algebraic problem. Since 
\begin{eqnarray}
\eta_-=\ln\left({\tan\theta_1/2\over\tan\theta_2/2}\right), \hskip1cm  \label{34}\\
\theta_1={1\over 2}(\theta_+-\theta_-),  \quad \theta_2={1\over 2}(\theta_++\theta_-),    \label{35}
\end{eqnarray}
we can solve for $\theta_-$ in terms of $\eta_-$ and $\theta_+$. For $\bar\theta$ not far from $\pi/2$, $\theta_-$ behaves essentially like $\eta_-$. Substituting the exact dependence of $\theta_-$ on $\eta_-$ and $\theta_+$ into $\tilde C_2(\theta_+, \theta_-, \phi_-)$ in Eq.\ (\ref{33}) and then averge over $\bar\theta$, we obtain
\begin{eqnarray}
\tilde A(\eta_-, \phi_-)={1\over 2\epsilon}\int_{\pi/2-\epsilon}^{\pi/2+\epsilon} d\bar\theta\, \tilde H(\eta_-, \phi_-,\bar\theta).   \label{36}
\end{eqnarray}
This is our main result, except for the 3D display of $\tilde A(\eta_-,\phi_-)$ after numerical computation. The essence of autocorrelation is basically already contained in Fig.\ 1 (b), where $\tilde H(\chi)$ is shown for various values of the jet cone width $\sigma$. For every value of $\chi$ there is a set of values of $\eta_-, \phi_-$ and $\bar\theta$, whose contour plot gives a representation of $\tilde A(\eta_-, \phi_-)$.

In Fig.\ 2 we plot $\tilde A(\eta_-, \phi_-)$ for $\sigma=0.2$. There is very little sensitivity on the dependence on $\epsilon$ in Eq.\ (\ref{36}); we set it at $\epsilon=0.3$. The dependence on $\sigma$ is, however, significant, as we have already seen in Fig.\ 1. The general shape of $\tilde A(\eta_-, \phi_-)$ is similar, the peak being broader for higher values of $\sigma$. The utility of this result is, of course, the reverse. When the data on $A(\eta_-, \phi_-)$ become available, we can use $A(\eta_-, \phi_-)$  to infer what the corresponding cone width $\sigma$ should be. To facilitate that deduction, we show in Fig.\ 3 (a) $\tilde A(\eta_-, 0)$  and (b) $\tilde A(0, \phi_-)$  for various values of $\sigma$. We see that the width in $\eta_-$ is numerically larger than that in $\phi_-$. 
While in the small  $\eta_-$ region one has  $\eta_-\sim \phi_-$, the two variables should not be directly compared since $\eta_-$ is not an angle. The more important implication of this result is that we have in Fig.\ 3 peaks in the measurable variables $\eta_-$ and $\phi_-$ for four values of the theoretical variable $\sigma$ whose magnitude is not known from first principles. Thus any experimental information on the autocorrelation can give us  direct information on the nature of the jet cone. As already mentioned in Sec.\ I, no trigger-related background subtraction is needed for autocorrelation, which is defined in terms of $C_2(1,2)$. Indeed, experimental data on autocorrelation are already available at low $p_T$ \cite{ja2,ja3}.

Finally, we note that the width of the peak in $\tilde A(\eta_-,0)$ is wider than that in $\tilde A(0,\phi_-)$, as shown in Fig.\ 3, not because of any flow effect in the longitudinal direction, such as in the finding in \cite{asw}. We have not considered any medium flow and its interaction with the shower partons. Our result follows strictly from a Gaussian distribution for both $\psi_1$ and $\psi_2$ of the shower partons relative to the jet axis. The variables $\theta_-$ and $\phi_-$ are differences of angles of $\vec p_1$ and $\vec p_2$ relative to the beam axis, and should not be mistaken to be the angles defined for an associated particle with respect to a trigger particle. 
The example given at the end of the preceding section on the angle between parton momenta can be translated now to the angle between detected hadrons, thereby illustrating the difference between autocorrelation and the correlation between a trigger and its associated particle.

\section{Conclusion}

We have determined the autocorrelation distribution in $\eta_-$ and $\phi_-$ for two pions produced in HIC in the intermediate $p_T$ region. We have used the parton recombination model to relate the hadronic angles to the partonic angles. By emphasizing thermal-shower parton recombination, we have exploited the equivalence of the angle between the two shower partons and that between the two observed pions. Thus when the data on autocorrelation at intermediate $p_T$ become available, the width of the observed peak can then be related to the width of the jet cone, which is a property of the jet physics in HIC that is the goal of this study.

It should be mentioned that our investigation has been focused on the angular relationship among the partonic and hadronic momenta. That relationship is independent of $p_T$ so long as all the transverse momenta in the problem are in the intermediate $p_T$ region. Since the thermal partons are soft, the shower parton and the pion that is formed by $TS$ recombination are roughly of the same magnitudes and are collinear. Thus the angle $\chi$ between two shower partons in a jet is the same as the angle $\chi$ between the two pions, independent of their momentum magnitudes. However, we have made use of the simplifying assumption that the jet cone has a width $\sigma$ that is independent of the momentum fraction of the shower partons. While it is sensible to make that assumption in this first attempt to calculate the autocorrelation, we expect the realistic situation to be more complicated. Once that dependence on momentum fraction is considered, the magnitude of the hard parton momentum becomes relevant, and the final result on autocorrelation will exhibit dependence on the $p_T$ range, even within the intermediate $p_T$ region $3<p_T<8$ GeV/c. Needless to add, when $p_T$ goes outside that region $TS$ recombination no longer dominates and the basis for our study in this paper will have to be revised.

Despite the simplifying assumption made in this paper, it will be of great interest to compare our result with the forthcoming data on autocorrelation at intermediate $p_T$. As far as we know, there exists no other theoretical study in the subject that relates the observables to the partonic structure in a jet, since no other hadronization scheme has been shown to be reliable in the $p_T$ range considered. In addition to pions one can also study proton and other heavier particles in the jets and their autocorrelations. Even in the pion sector alone it is of interest to consider the complications arising from different charge states, since multiple shower partons in a jet can exhibit dependence on their flavors. Thus there remains much to be learned in the subject, in which the present study is only a beginning.

\section*{Acknowledgment}
We are grateful to  T.\ Trainor for valuable criticisms of our draft manuscript and to  L.\ Ray for helpful discussions.    This work was supported, in part,  by the U.\ S.\ Department of Energy under
Grant No. DE-FG03-96ER40972.

\vskip1cm
\begin{center}
\section*{Figure Captions}
\end{center}

\begin{description}
\item
Fig.\ 1. (color online) (a) Distributions in the angle $\chi$ for four values of the width parameter $\sigma$ of the jet cone; (b) normalized distributions $\tilde H(\chi)$.

\item
Fig.\ 2. (color online) Normalized autocorrelation function $\tilde A(\eta_-, \phi_-)$  in 3D plot in  $\eta_-$-$\phi_-$ for $\sigma=0.2$.

\item
Fig.\ 3. (color online) (a) Normalized autocorrelation $\tilde A(\eta_-, 0)$ for four values of $\sigma$; (b) normalized $\tilde A(0, \phi_-)$.

\end{description}

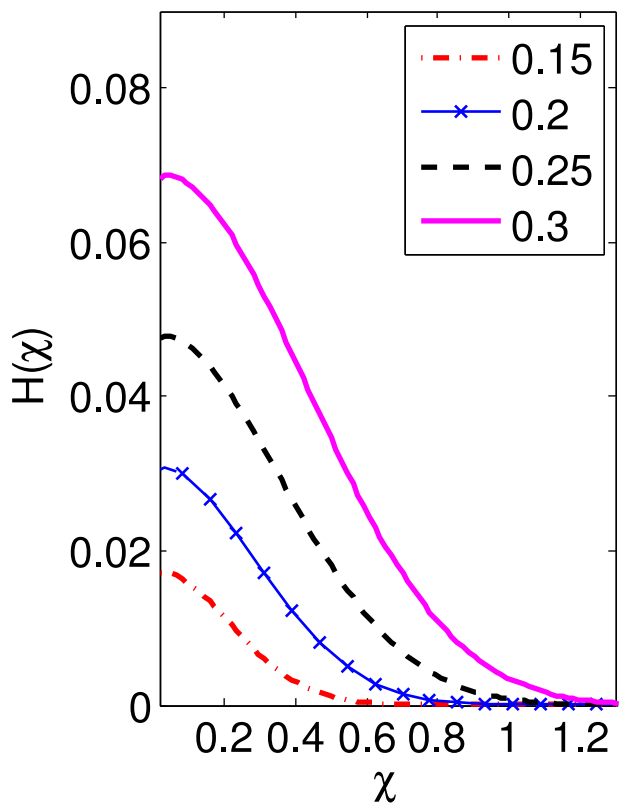
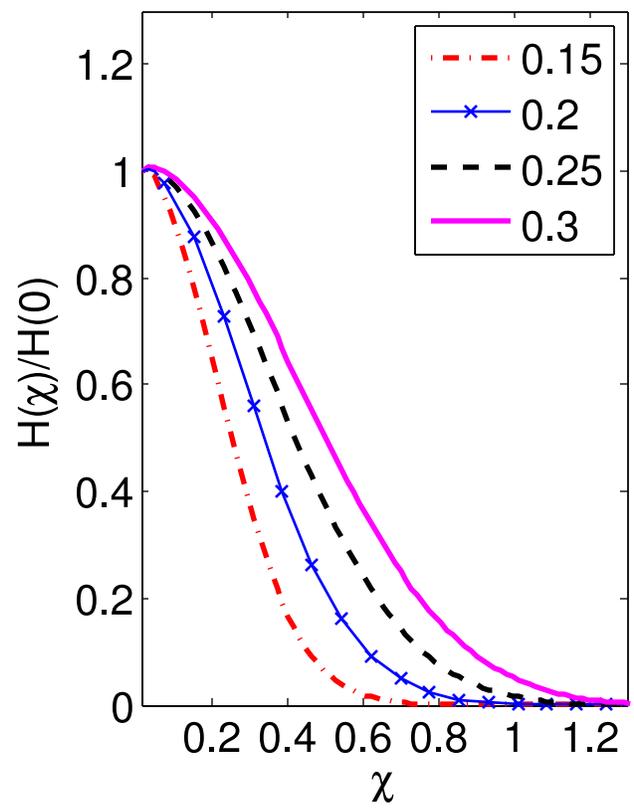
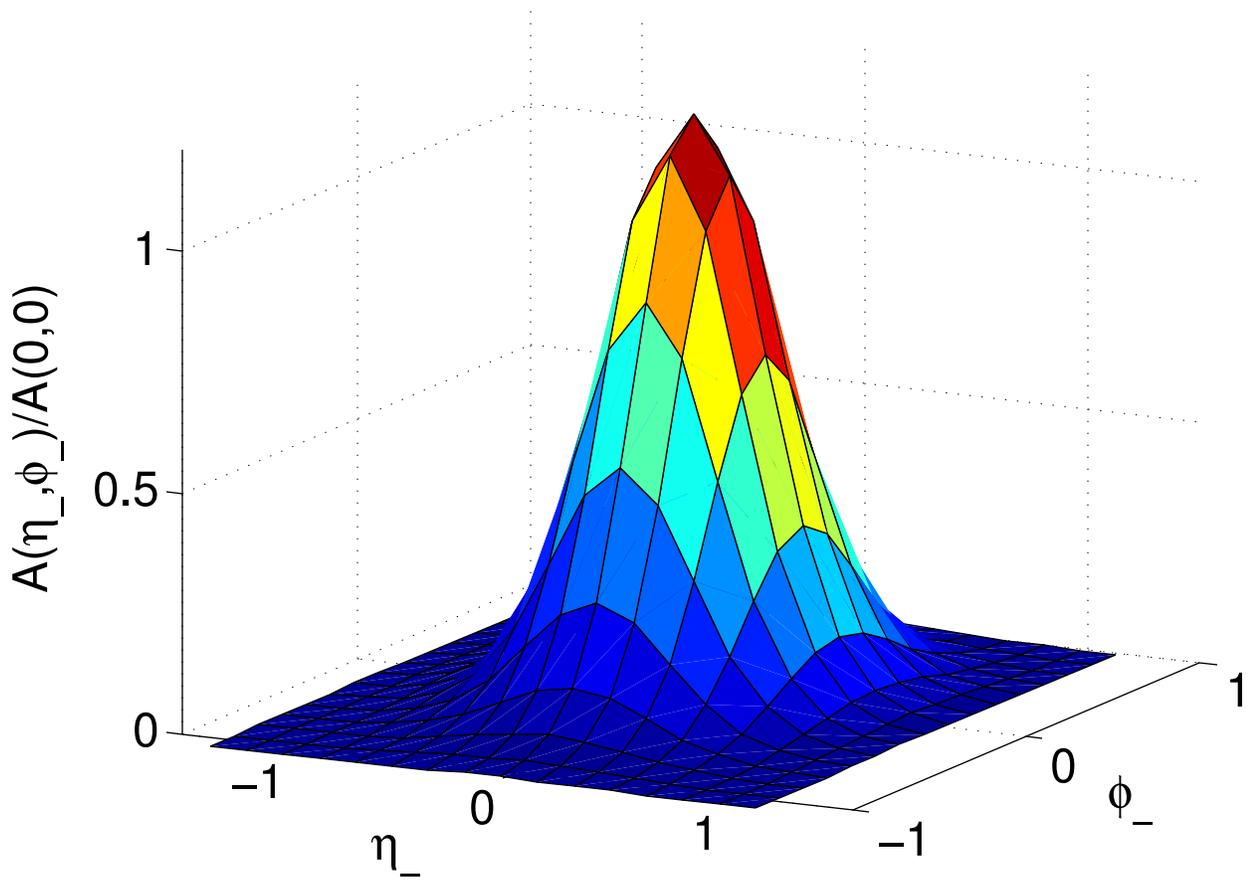
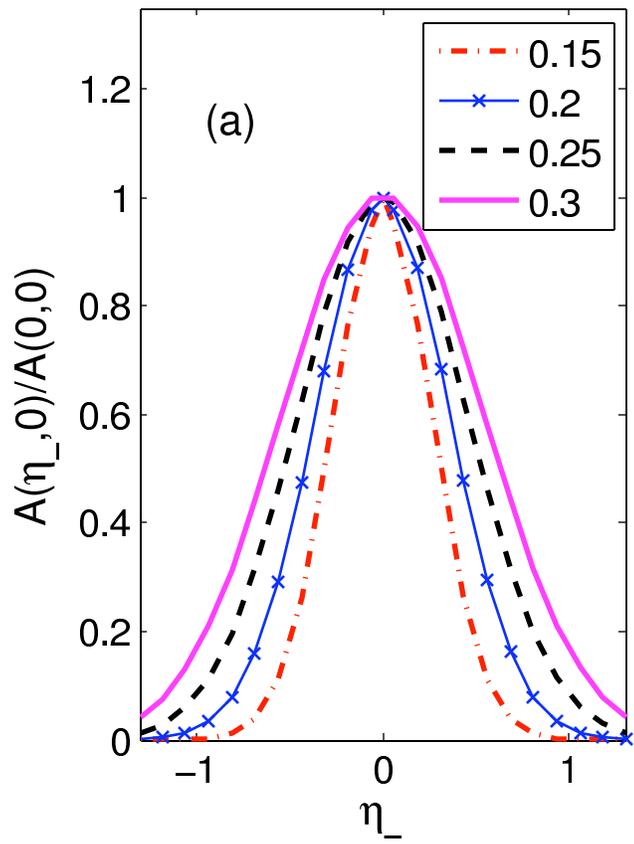
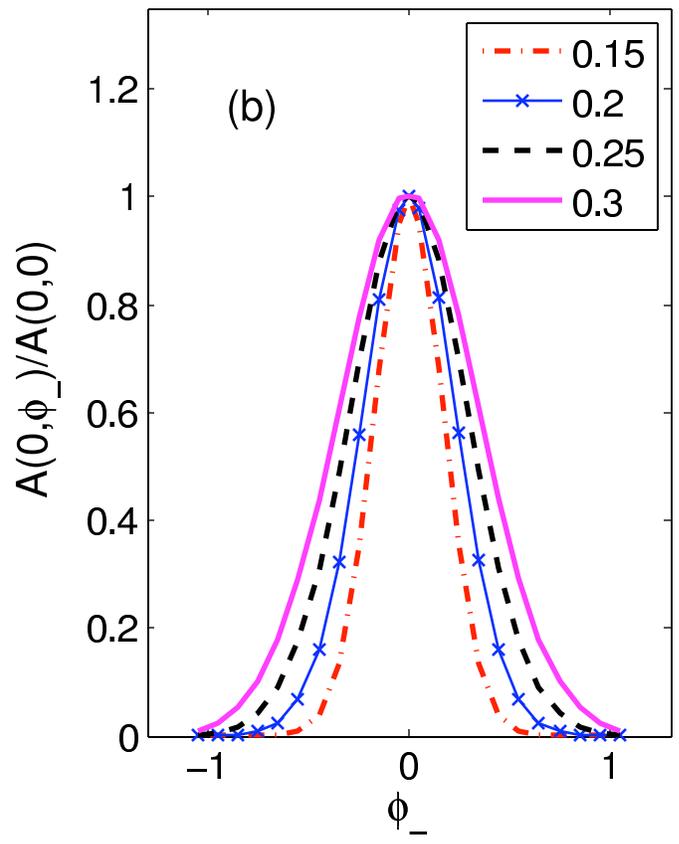
\end{document}